\documentclass{article}
\usepackage{spconf,amsmath,graphicx}
\usepackage{comment}
\usepackage{url}
\usepackage{booktabs}
\usepackage{amssymb}

\usepackage{enumitem}
\setlist{nosep, leftmargin=14pt}

\usepackage{mwe} 

\title{Automatic segmentation of lung findings in CT and application to long COVID}
\name{
\begin{tabular}{@{}c@{}}
Diedre S. Carmo$^{\star\dagger}$ \qquad Rosarie A. Tudas$^{\ddagger}$ \qquad Alejandro P. Comellas$^{\ddagger}$ \\ \qquad Leticia Rittner$^{\star}$ \qquad Roberto A. Lotufo$^{\star}$ \qquad Joseph M. Reinhardt$^{\dagger}$ \qquad Sarah E. Gerard$^{\dagger}$
\end{tabular}}
\address{$^{\star}$ MICLab, School of Electrical and Computing Engineering, University of Campinas, Campinas, Brazil\\
         $^{\dagger}$ Roy J. Carver Department of Biomedical Engineering, University of Iowa, Iowa City, US\\
         $^{\ddagger}$ Department of Internal Medicine, Carver College of Medicine, University of Iowa, Iowa City, US\\\\\small Versão em português brasileiro submetida para o XV EADCA 2023}

\begin{document}
%
\maketitle
\begin{abstract}

Automated segmentation of lung abnormalities in computed tomography is an important step for diagnosing and characterizing lung disease. In this work, we improve upon a previous method and propose S-MEDSeg, a deep learning based approach for accurate segmentation of lung lesions in chest CT images. S-MEDSeg combines a pre-trained EfficientNet backbone, bidirectional feature pyramid network, and modern network advancements to achieve improved segmentation performance. A comprehensive ablation study was performed to evaluate the contribution of the proposed network modifications. The results demonstrate modifications introduced in S-MEDSeg significantly improves segmentation performance compared to the baseline approach. The proposed method is applied to an independent dataset of long COVID inpatients to study the effect of post-acute infection vaccination on extent of lung findings. Open-source code, graphical user interface and pip package are available at \url{https://github.com/MICLab-Unicamp/medseg}.

\end{abstract}
\section{Introduction}
\label{sec:intro}


It is estimated that 43\% of those infected with COVID-19 suffer from symptons months after the acute disease period, a condition called ``long COVID''~\cite{chen2022global}. Reported symptoms of long COVID include fatigue, shortness of breath, neurological problems~\cite{beghi2022short}, and lung findings on computed tomography (CT)~\cite{cho2022quantitative}. The effectiveness of vaccination in improving outcomes for the acute infection has been demonstrated. However, the effects of post-acute vaccination in long COVID are not clear, with some evidence suggesting vaccination reduces the risk of developing long COVID~\cite{notarte2022impact}.



CT imaging can provide valuable information for studying long COVID as it reveals the extent and spatial distribution of lung findings such as consolidation and ground glass opacities (GGO), helping characterize the disease and its severity~\cite{wasilewski2020covid}. Manual annotation by radiologists is considered the gold standard for detecting abnormalities in chest CT. However, with large datasets of high-resolution CT scans, this process is time consuming.

To facilitate efficient segmentation of lung findings, not only for COVID-19 but other diseases such as lung cancer, many automated deep learning based segmentation methods have been proposed~\cite{shi2020review}, usually supervised by manual annotation. Lung findings commonly have fuzzy boundaries which introduces subjectivity in the manual labeling~\cite{fan2022gfnet}. This tends to keep achieved Dice coefficients~\cite{sudre2017generalised} around the 0.6-0.8 range with significant standard deviation on different datasets. However, the manual segmentations provide enough information to train a neural network to achieve satisfactory segmentation results~\cite{carmo2021rapidly} for the purpose of computing infection radiomics~\cite{zhang2021deep}. There have been reports of comparable performance in detection of lesions when compared to radiologists~\cite{ni2020deep}, and enhanced outcome prediction when using both radiomic features derived from segmentations and clinical data~\cite{xudong2022artificial}. Therefore, a pre-trained deep learning segmentation network can be used to quickly compute segmentations and radiomics of large CT datasets, such as the percentage of lung involvement, to better understand the disease behaviour.

In previous work, we proposed CoEDet for segmentation of lung and abnormalities in COVID-19 patients, which demonstrated top performance~\cite{carmo2021multitasking, carmo2021rapidly}. In this work, we propose Specialized Modified EfficientDet segmentation (S-MEDSeg) which builds upon CoEDet by incorporating recent known and novel network advancements to achieve robust segmentation of lung lesions in CT images. Additionally, we analyze the lobar distribution of findings in an independent cohort of long COVID inpatients~\cite{cho2022quantitative} using an existing lobar segmentation approach~\cite{gerard2019pulmonary}. Finally, we study the association of CT-based lung findings severity with patient demographics and vaccination status on the same cohort. This is the first application of deep learning segmentation models in reporting the lobar distribution of findings and their association with vaccination status in long COVID patients.

\section{Materials and Methods}
\label{sec:mat_method}


\subsection{Datasets}
\subsubsection{Training and Validation}
Datasets with lung and findings manual annotation from the following sources were used in the study: IdeiaGov, CoronaCases, MSC, MOSMED and MICCAI Challenge, which were the same development datasets used in CoEDet~\cite{carmo2021multitasking}. In total, 490 volumetric CT images with a combined 78531 512x512 axial slices were utilized, composed of a concatenation of training (80\%) and validation (20\%) splits from all datasets. The different data sources provided variations in imaging and reconstruction protocols. 

\subsubsection{Long COVID}
A dataset of 88 long COVID patients was used for independent qualitative evaluation and analysis. Data was collected at the University of Iowa Hospitals and Clinics~\cite{cho2022quantitative}. Adults with history of COVID-19 infection confirmed by a positive antigen or reverse transcriptase-polymerase chain reaction and remained symptomatic 30 days or more following diagnosis were prospectively enrolled between June 2020 and April 2022. Patient data collected included age, sex, days between COVID-19 diagnosis and CT exam, and vaccination status. Vaccination status was defined as receiving at least 1 dose of the COVID mRNA vaccine after acute infection diagnosis and before the CT scan, yielding 55 patients classified as post-infection vaccinated and 33 patients classified as unvaccinated. Only patients who received a CT scan and required inpatient care during acute infection (i.e. hospitalized or ICU) were included in this study.




\subsection{S-MEDSeg}
\label{ssec:seg_method}
CoEDet is a fully convolutional neural network (FCNN) that outputs voxelwise segmentations for chest CT images. It is a 2.5D network as the input contains three neighboring axial slices. The main idea behind CoEDet is using features from an EfficientNet~\cite{tan2019efficientnet} backbone pre-trained on ImageNet. They are fed into a bidirectional feature pyramid network (BiFPN) that generates multi-resolution output features that can be used for downstream tasks. We have specifically optimized this EfficientDet~\cite{tan2020efficientdet} inspired architecture before for lung and findings segmentation in COVID-19~\cite{carmo2021multitasking} and airway segmentation~\cite{carmo2022open}.

\begin{figure*}
\centerline{\includegraphics[width=\textwidth]{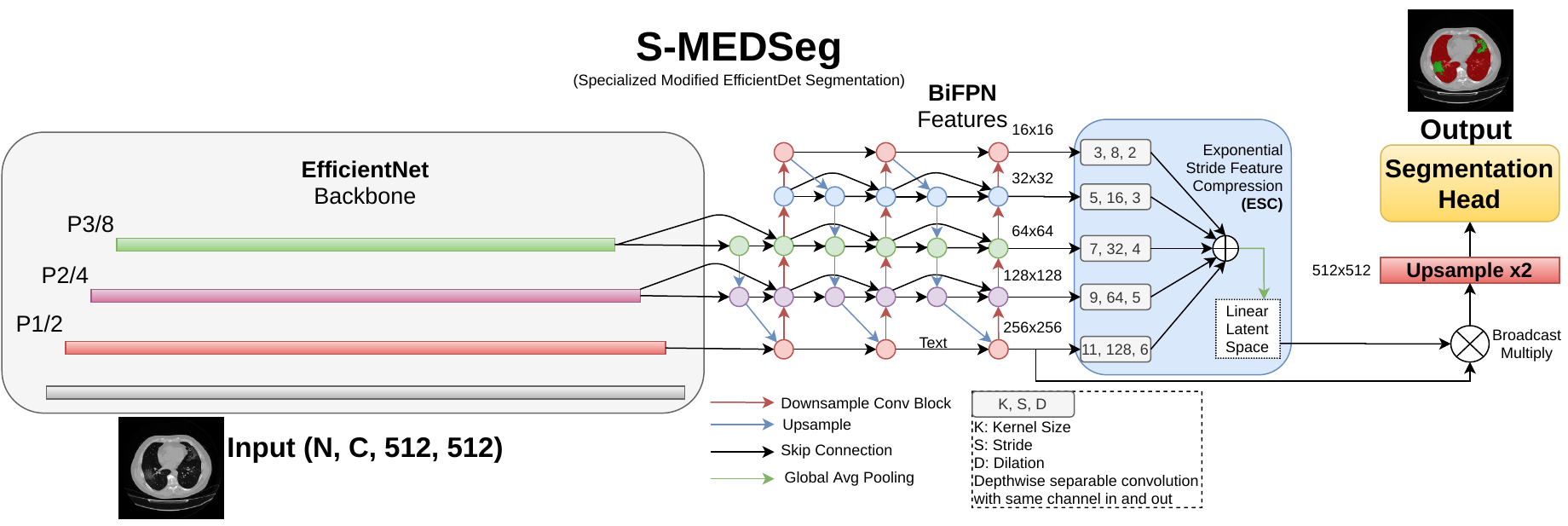}}
\caption{\label{fig:method_esc}Schematic of S-MEDSeg architecture. Features from an EfficientNet backbone are input to a feature pyramid network. ESC linear embedding derived from those features are used to weight high resolution features. The segmentation head contains three blocks of depth-wise separable convolutions, batch norm, and swish activation.}
\end{figure*}

S-MEDSeg improves upon CoEDet using a novel strategy to incorporate all features from the BiFPN output, called exponential stride compression (ESC). ESC transforms features into a latent space that weights the contribution of each channel before passing features to the segmentation head. ESC uses large dilated depthwise separable convolution kernels with exponentially increasing stride (Figure~\ref{fig:method_esc}). Additional modifications we tested when designing S-MEDSeg were inspired in part by the recent ConvNext work~\cite{liu2022convnet} and are described next. Transposed convolutional layers were replaced with simple bilinear upsampling. The ConvNext~\cite{liu2022convnet} backbone was tested in place of EfficientNet. For training, we tested increasing the patch size, increasing the batch size, and extracting patches only from slices containing annotation. For optimization we tested AdamW~\cite{loshchilov2018decoupled} with weight decay as well as using exponential learning rate decay.


\subsection{Evaluation and Experiments}
An ablation study was used to study the contribution of the different modifications incorporated in S-MEDSeg. The previous CoEDet~\cite{carmo2021multitasking} architecture, which has performance superior to 2D, 2.5D and 3D UNet in our dataset, was used as a baseline for quantitative evaluation for the proposed improvements. The Dice coefficient on the holdout validation split was used for evaluation. A Wilcoxon rank sum test~\cite{carmo2023automated} was used to statistically compare network performance after all modifications. We did not notice a significant difference in early experiments from using windowing and intensity normalization to using original HU values. Networks were trained with random patches, and an early stopping patience of 100 epochs. The weights with the lowest validation loss were used for evaluation. A NVIDIA A100 or 2080 Ti GPU was used for training.

Qualitative evaluation was performed on the long COVID-19 dataset. Additionally, as a quality control lung segmentations produced by S-MEDSeg were compared to lungmask~\cite{hofmanninger2020automatic}. Lobe segmentations for the long COVID dataset were generated using the top performing method in the LOLA11 grand challenge~\cite{gerard2019pulmonary,gerard2021ct}. Percentage of involvement (POI) were computed for whole lung and per lobe. A multivariate linear regression was performed to study the association between total POI and vaccination status, controlling for age, sex, and days between CT scan and diagnosis.

\section{Results}
\label{sec:results}

Results from the ablation study are reported in Table~\ref{tab:results}. S-MEDSeg performance was significantly better compared to the CoEDet baseline ($p<0.05$) with higher mean and lower standard deviation Dice. S-MEDSeg has 17\% lower false positive error rate and 4\% lower false negative error rate. In addition, S-MEDSeg also surpassed nnUNet~\cite{isensee2021nnu} in a 5-fold evaluation, achieving $0.6737\pm0.0261$ Dice against nnUNet's $0.5829\pm0.0270$.



\begin{figure*}[!h]
\centerline{\includegraphics[width=\textwidth]{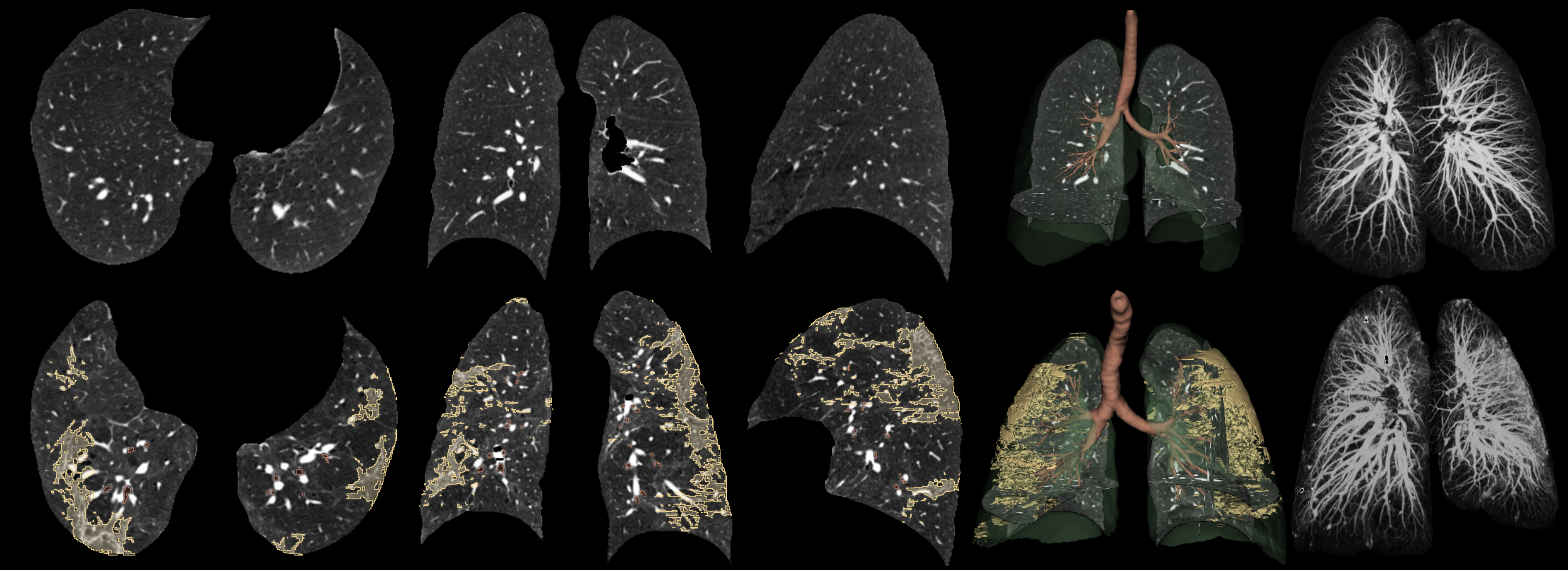}}
\caption{\label{fig:intro}Representative S-MEDSeg results for subjects with 0\%POI (top row) and 14\% POI (bottom row). First three columns correspond to CT slices with findings overlaid in yellow for axial, coronal, and sagittal views. Fourth column shows renderings of segmentation results including lung (green), airway (brown), and findings (yellow). Fifth column shows maximum intensity projection.}
\end{figure*}
\begin{figure}[!h]
\centerline{\includegraphics[width=\columnwidth]{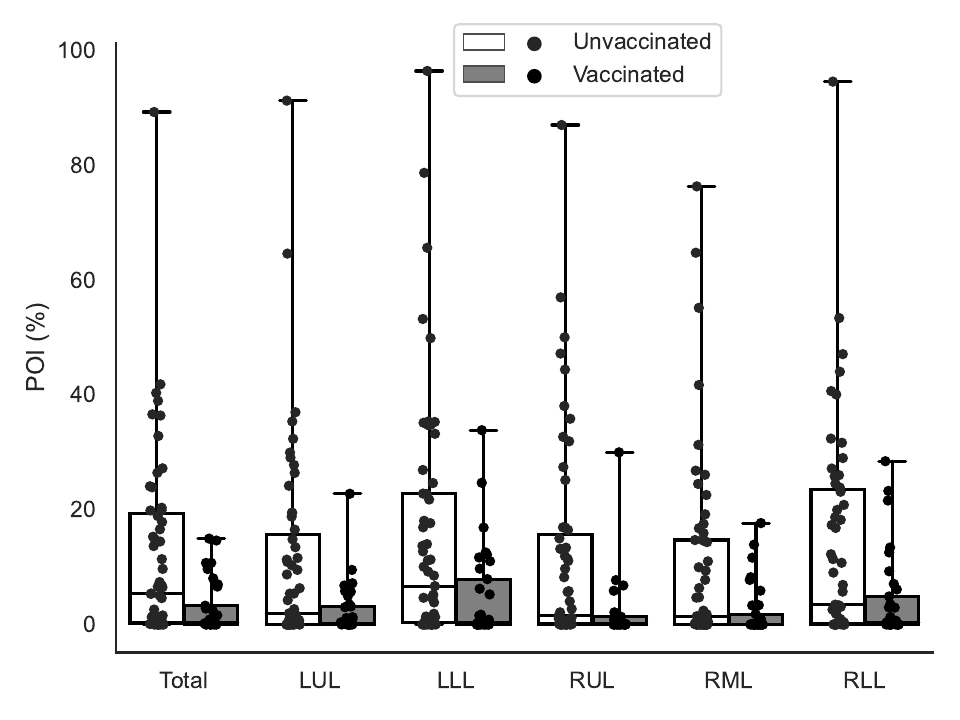}}
\caption{\label{fig:boxplot} Percentage of involvement (POI) for total lung, left upper lobe (LUL), left lower lobe (LLL), right upper lobe (RUL), right middle lobe (RML), and right lower lobe (RLL). Subjects are stratified by vaccination status and bars extend to min and max values.}
\end{figure}

Segmentation results for the long COVID dataset were found to be of satisfactory quality by visual inspection. Representative results are visualized in Fig.~\ref{fig:intro}. There was 95\% Dice agreement between lung segmentations produced by S-MEDSeg and lungmask. Lobar analysis revealed the lower lobes had more involvement (Fig.~\ref{fig:boxplot}).
A t-test revealed subjects with post infection vaccination had a decrease in POI compared to unvaccinated subjects ($p=0.002$). However, the multivariate regression found that the association of POI and vaccination after initial infection was not significant ($p=0.07$) after correction for age, sex, and days between diagnosis and CT. 



\begin{table}[!h]
\centering
\resizebox{\columnwidth}{!}{
\begin{tabular}{cccccccccc}
\\\toprule
\textbf{ESC} & \textbf{IPS}          & \textbf{IBS}     & \textbf{SWA}         & \textbf{WLRD}       & \textbf{CNF} & \textbf{UPS}    & \textbf{AW}      & \textbf{3D Dice}             \\\midrule
             &                       &                  &                      &                     &              &                 &                  & $0.6023\pm0.2242$   \\
             &                       &                  &                      & \checkmark          &              &                 &                  & $0.6280\pm0.2328$   \\
             & \checkmark            &                  & \checkmark           & \checkmark          & \checkmark   &                 & \checkmark       & $0.6413\pm0.2081$   \\
             & \checkmark            &                  &                      &                     &              &                 &                  & $0.6535\pm0.2067$   \\
             & \checkmark            &                  & \checkmark           &                     &              &                 &                  & $0.6577\pm0.2114$   \\
             & \checkmark            &                  & \checkmark           & \checkmark          &              &                 & \checkmark       & $0.6649\pm0.1740$   \\
             & \checkmark            &                  & \checkmark           &                     &              &                 &                  & $0.6657\pm0.1952$   \\
             & \checkmark            &                  & \checkmark           & \checkmark          &              &                 &                  & $0.6742\pm0.1891$   \\
             & \checkmark            &                  & \checkmark           & \checkmark          &              &                 & \checkmark       & $0.6759\pm0.2060$   \\
             & \checkmark            &                  & \checkmark           & \checkmark          &              & \checkmark      & \checkmark       & $0.6761\pm0.1986$   \\
             & \checkmark            & \checkmark       & \checkmark           & \checkmark          &              & \checkmark      & \checkmark       & $0.6842\pm0.1840$   \\ 
\checkmark   & \checkmark            & \checkmark       & \checkmark           & \checkmark          &              & \checkmark      & \checkmark       & $0.6892\pm0.1832$   \\\midrule 
\end{tabular}
}
\caption{\label{tab:results}Ablation study results showing Dice coefficient for various methodology modifications. The first row is a baseline CoEDet architecture and the last row corresponds to the proposed MEDSeg. Studied modifications include: Exponential Stride Feature Compression (ESC), increased patch size (IPS) and batch size (IBS) to 256x256 and 60 respectively, inclusion only of Slices With Annotation (SWA), 1e-5 weight and 0.985 exponential learning rate decay (WLRD), replacing: EfficientNet with ConvNext features (CNF); transposed convolutions with bilinear upsample (UPS); and using the AdamW optimizer (AW) with 1e-4 initial learning rate.}
\end{table}

\section{Discussion}
\label{sec:discussion}
Results indicate the proposed S-MEDSeg method improves upon the CoEDet baseline. We found that replacing learned transposed convolution layers with simple bilinear feature upsampling improved results. Additionally, we found the recently proposed ConvNext backbone did not perform better than EfficientNet. We speculate that this is due to the loss of spatial information in the lower resolution ConvNext features compared to EfficientNet. Increasing patch size, batch size, and learning only from slices with annotation had positive impacts on performance. Our proposed supervision of segmentation features (ESC) through a latent space derived from BiFPN features also improved results. These results indicate in general that less upsampling is beneficial for semantic segmentation based on FPNs, and that one should attempt to make use of all features provided by a FPN. Moreover, as found recently by ConvNext~\cite{liu2022convnet}, high kernel size convolution should not be overlooked in CNN architectural design. 

A t-test revealed that the unvaccinated group had significantly more findings compared to the post acute infection vaccinated group, however, after controlling for other variables in a multivariate regression the association was not significant. Our analysis indicated there was more involvement of the lower lobes which is consistent with previous literature~\cite{gerard2021ct,xudong2022artificial}. 
The main limitation of this study is the small sample size for long COVID inpatients.


\section{Conclusion}
\label{sec:conclusion}
The proposed S-MEDSeg architecture significantly improved segmentation of lung findings compared to a baseline approach. The proposed method was deployed in an independent dataset of long COVID subjects who were hospitalized during acute infection, to investigate lobar distribution of lesions and the relationship between vaccination status and the extent of findings. Post-infection vaccinated subjects had fewer findings compared to unvaccinated subjects, however, after correction for confounding variables this result was not statistically significant. Future work will use a larger sample size and study other factors that might influence the severity of long COVID in chest CT. Furthermore, we will explore extending the S-MEDSeg framework to the segmentation of other pulmonary structures and lesions, expanding it to general chest CT segmentation instead of specialization to specific chest targets.

\section{Compliance with ethical standards}
\label{sec:ethics}
This study was performed in line with the principles of the Declaration of Helsinki. For included data that is not of public domain, study protocols were approved by the institutional review board and were Health Insurance Portability and Protection Act–compliant. Participants provided written informed consent prior to inclusion. The RadVid19 initiative was approved by the Brazilian National Research Ethics Commission. 


\section{Acknowledgments}
\label{sec:acknowledgments}

Part of data used on this work comes from a challenge run by the RadVid19 program, made possible thanks to the public notice published by the IdeiaGov initiative of the Economic Development Secretary of the State of São Paulo, and is related to the public bidding no. 02/2020, process  SDE-PRC-2020/00148. Diedre Carmo thanks the support from grants 2019/21964-4 and 2022/02344-8, São Paulo Research Foundation (FAPESP). R Lotufo is partially supported by CNPq (The Brazilian National Council for Scientific and Technological Development) under grant 310828/2018-0. J Reinhardt is a shareholder in VIDA Diagnostics, Inc.

\bibliographystyle{ieetr.bst}
\bibliography{isbi}

\end{document}